\documentclass[graybox]{svmult}

\usepackage{amssymb}
\usepackage{amsmath}

\usepackage{indentfirst}
\usepackage{xspace} 

\usepackage[english]{babel}
\usepackage[latin1]{inputenc}

\usepackage{mathptmx}       
\usepackage{helvet}         
\usepackage{courier}        
\usepackage{type1cm}        
                            
\usepackage{makeidx}          
\usepackage{graphicx}         
\usepackage{multicol}         
\usepackage[bottom]{footmisc} 

\makeindex             

\usepackage[usenames,dvipsnames]{pstricks}
\usepackage{epsfig}
\usepackage{pst-grad} 
\usepackage{pst-plot} 
                                 
\renewcommand{\div}{\nabla\cdot}
\newcommand{\Ma}{\mathrm{Ma\,}}
\newcommand{\Fr}{\mathrm{Fr}}
\newcommand{\St}{\mathrm{St\:}}
\newcommand{\Ve}{{V_\epsilon}}
\newcommand{\R}{\mathbb{R}}
\newcommand{\Rr}{\mathbf{R}}
\newcommand{\B}{\mathbf{B}}
\newcommand{\A}{\mathbf{A}}
\newcommand{\J}{\mathbf{J}}
\renewcommand{\P}{\mathbf{P}}
\newcommand{\T}{\mathbf{T}}
\renewcommand{\S}{\mathbf{S}}
\renewcommand{\D}{\mathbb{D}}
\newcommand{\Dv}{\mathbf{D}}
\renewcommand{\O}{\mathrm{O}}
\newcommand{\Id}{\mathbf{I}}
\newcommand{\g}{\vec{g}}
\newcommand{\x}{\vec{x}}
\renewcommand{\u}{\vec{u}}
\renewcommand{\phi}{\varphi}
\newcommand{\ttau}{\boldsymbol{\tau}}

\newcommand{\dt}{\partial_t}
\newcommand{\od}[2]{\frac{d #1}{d #2}}
\newcommand{\pd}[2]{\frac{\partial #1}{\partial #2}}
\newcommand{\tr}{\mathop{\mathrm{tr}}}
\newcommand{\Mat}{\mathop{\mathrm{Mat}}}
\newcommand{\range}{\mathop{\mathrm{range}}}

\begin{document}

\title{Two-fluid barotropic models for powder-snow avalanche flows}
\titlerunning{Two-fluid models for powder-snow avalanche flows}
\author{Yannick Meyapin, Denys Dutykh and Marguerite Gisclon}
\authorrunning{Y. Meyapin, D. Dutykh \& M. Gisclon}

\institute{Yannick Meyapin \at LAMA, UMR 5127 CNRS, Universit\'e de Savoie, 73376 Le Bourget-du-Lac Cedex, France, \email{Yannick.Meyapin@etu.univ-savoie.fr} \and
Denys Dutykh \at LAMA, UMR 5127 CNRS, Universit\'e de Savoie, 73376 Le Bourget-du-Lac Cedex, France, \email{Denys.Dutykh@univ-savoie.fr} \and
Marguerite Gisclon \at LAMA, UMR 5127 CNRS, Universit\'e de Savoie, 73376 Le Bourget-du-Lac Cedex, France, \email{Marguerite.Gisclon@univ-savoie.fr}}

\maketitle

\abstract*{In the present study we discuss several modeling issues of powder-snow avalanche flows. We take a two-fluid modeling paradigm. For the sake of simplicity, we will restrict our attention to barotropic equations. We begin the exposition by a compressible model with two velocities for each fluid. However, this model may become non-hyperbolic and thus, represents serious challenges for numerical methods. To overcome these issues, we derive a single velocity model as a result of a relaxation process. This model can be easily shown to be hyperbolic for any reasonable equation of state. Finally, an incompressible limit of this model is derived.}

\abstract{In the present study we discuss several modeling issues of powder-snow avalanche flows. We take a two-fluid modeling paradigm. For the sake of simplicity, we will restrict our attention to barotropic equations. We begin the exposition by a compressible model with two velocities for each fluid. However, this model may become non-hyperbolic and thus, represents serious challenges for numerical methods. To overcome these issues, we derive a single velocity model as a result of a relaxation process. This model can be easily shown to be hyperbolic for any reasonable equation of state. Finally, an incompressible limit of this model is derived.}

\section{Introduction}

Snow avalanches represent a serious problem for society in mountain regions. The avalanche winter of 1999 attracted a lot of attention to this hazardous natural phenomenon \cite{Ancey2006, Lied2006}. Further development of mountain regions requires an adequate level of avalanche safety. Therefore, avalanche protective measures (deflecting and catching dams) become increasingly important \cite{Johannesson2009}. During the same winter, several avalanches overran avalanche dams, underlining the need for further research in this field. Proper design of protecting structures necessitates profound understanding of the snow avalanches flow and of the interaction process with dams and other obstacles \cite{Dutykh2009, Naaim-Bouvet2002}.

Natural snow avalanches are believed to consist of three different layers: a dense core, a fluidised layer and a suspension cloud. Sometimes the surrounding powder cloud is absent and we speak about an avalanche in the flowing r\'egime. Obviously, transition boundaries between these layers are not sharp and this classification is rather conventional.

The dense core consists of snow particles in persistent frictional contact \cite{Issler2003}. The density is of the order of 300 kg/m$^3$ and the depth of this layer does not exceed 3 m. The fluidised r\'egime is characterized by particle's mean-free-paths up to several particle's diameters. This dynamics at microscopic level explains more fluid-like behaviour at large scales. The density of this layer is in the range of 50 - 100 kg/m$^3$ and the height is about 3 - 5 m. To model successfully this kind of flows it is crucial to know the complex fluid rheology. Finally, these two interior layers can be covered by the powder cloud which is a turbulent suspension of snow particles in the air. The density ranges from 4 to 20 kg/m$^3$ and an avalanche in aerosol r\'egime can reach the height of 100 m or more \cite{Rastello2004}. This flow is driven essentially by turbulent advection and particles collisions are unimportant.

In the present study we are concerned with some questions of powder-snow avalanche modelling. Since the interface cannot be defined for this type of flows, we choose the modelling paradigm of two-phase flows. In this approach the governing equations of each phase are spatially averaged to come up with the description of the fluid mixture \cite{Ishii1975, Rovarch2006}. 

It is known \cite{Rastello2004} that the front of such an avalanche can develop the speed\footnote{When we estimate the Mach number magnitude, the particle characteristic velocity should be taken. However, this information is not easily accessible and we took the maximum front velocity. It can lead to some overestimation of the Mach number.} $u_f \approx$ 100 m/s. For comparison, the speed of sound $c_0$ in the air is about 300 m/s. It means that the local Mach number $\Ma$ can reach the value of
\begin{equation*}
  \Ma := \frac{u_f}{c_0} \approx 0.33.
\end{equation*}
Hence, compressible effects may become important. That is why, we begin our exposition with a compressible model. Then, we gradually simplify it to come up with an incompressible one at the end of the present article. The goal is achieved by taking the limit as the Mach number tends to zero.

The present article is organized as follows. In Section \ref{sec:2phase} we present a barotropic compressible two-phase model with two velocities. Then, this model is simplified in Section \ref{sec:relax} using a velocity relaxation process. The incompressible limit of resulting system is derived in Section \ref{sec:lowMach}. Finally, several conclusions and perspectives are drawn out in Section \ref{sec:concl}.

\section{Two-phase flow modelling}\label{sec:2phase}

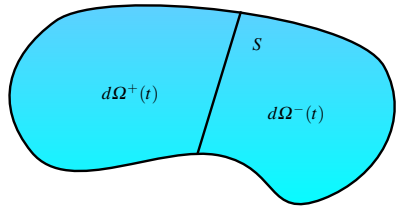
\begin{figure}[b]
  \begin{center}
	\scalebox{0.8} 
	{
	\begin{pspicture}(0,-1.9729664)(6.9945483,1.9723125)
	\definecolor{color1g}{rgb}{0.4,0.8,1.0}
	\psbezier[linewidth=0.04,fillstyle=gradient,gradlines=2000,gradbegin=color1g,gradmidpoint=1.0](3.7758446,-0.87319964)(4.7539034,-1.0815277)(4.5428023,-1.9529666)(5.4758444,-1.5931996)(6.408887,-1.2334328)(6.9745483,-0.27117625)(6.4958444,0.6068004)(6.017141,1.484777)(1.8465083,1.9523125)(1.0358446,1.3668003)(0.22518091,0.7812883)(0.0030853755,-0.038851038)(0.6358446,-0.81319964)(1.2686038,-1.5875483)(2.7977855,-0.66487163)(3.7758446,-0.87319964)
	\psline[linewidth=0.04cm](3.3958447,-0.8331996)(4.1158447,1.5268004)
	\usefont{T1}{ptm}{m}{n}
	\rput(2.2772508,0.15680036){$d\Omega^+ (t)$}
	\usefont{T1}{ptm}{m}{n}
	\rput(5.037251,-0.18319964){$d\Omega^- (t)$}
	\usefont{T1}{ptm}{m}{n}
	\rput(4.387251,0.9768004){$S$}
	\end{pspicture} 
	}
	\caption{An elementary fluid volume $d\Omega$ occupied by two phases.}
	\label{fig:volumefract}
	\end{center}
\end{figure}

Let us consider a domain $\Omega \subseteq \R^3$ where a simultaneous flow of two barotropic fluids occurs. All quantities related to the heavy and light fluids will be denoted by $+$ and $-$ correspondingly. In view of application to snow avalanches, one can consider the heavy fluid of being constituted of snow particles and the light fluid is the air. When the mixing process is extremely complicated and it is impossible to follow the interface between two fluids, the classical modelling procedure consists in applying a volume average operator \cite{Ishii1975, Rovarch2006}. Thereby, we make appear two additional variables $\alpha^\pm (\x, t)$, $\x\in\Omega$ which are called the volume fractions and defined as:
\begin{equation*}
  \alpha^\pm(\x,t) := \lim\limits_{\stackrel{|d\Omega|\to 0}{\x \in d\Omega}}
  \frac{|d\Omega^\pm|}{|d\Omega|},
\end{equation*}
the heavy fluid occupies volume $d\Omega^+\subseteq d\Omega$ and the light one the volume $d\Omega^-\subseteq d\Omega$ (see Figure \ref{fig:volumefract}) such that
\begin{equation}\label{eq:dOmega}
  |d\Omega| \equiv |d\Omega^+| + |d\Omega^-|.
\end{equation}
From the relation (\ref{eq:dOmega}) it is obvious that $\alpha^+(\x,t) + \alpha^-(\x,t) \equiv 1$, $\forall \x \in \Omega$.

After performing the averaging process, one obtains two equations of mass and momentum conservation:
\begin{eqnarray}\label{eq:mass4}
	\dt(\alpha^{\pm}\rho^{\pm}) + \div(\alpha^{\pm}\rho^{\pm}\u^{\pm}) &=& 0, \\
	\dt(\alpha^{\pm}\rho^{\pm}\u^{\pm}) + \div(\alpha^{\pm}\rho^{\pm}\u^{\pm}\otimes\u^{\pm}) + 
	\alpha^{\pm}\nabla p &=& \div(\alpha^{\pm}\ttau^{\pm})+\alpha^{\pm}\rho^{\pm}\g, \label{eq:momentum4}
\end{eqnarray}
where $\rho^\pm (\x,t), \u^\pm (\x,t), \ttau^\pm (\x,t)$ are densities, velocities and viscous stress tensors of each fluid respectively. Traditionally, the vector $\g$ denotes the gravity acceleration. We assume that both fluids share the same pressure\footnote{In general, this kind of assumptions is reasonable, since relaxation processes will tend to equilibrate the system when time evolves.} $p = p^\pm (\rho^\pm)$ and equations of state of each phase fulfill minimal thermodynamical requirements: 
\begin{equation}\label{eq:EOS}
  p^\pm(\rho^\pm) > 0, \quad \pd{p^\pm (\rho^\pm)}{\rho^\pm} > 0, \quad \mbox{for} \quad \rho^\pm > 0.
\end{equation}
In order to obtain a well-posed problem, governing equations (\ref{eq:mass4}), (\ref{eq:momentum4}) should be completed by appropriate initial and boundary conditions.

If we assume both fluids to be Newtonian, the viscous stress tensor $\ttau^\pm$ takes the following classical form:
\begin{equation}\label{eq:viscous}
  \ttau^\pm = \lambda^{\pm}\tr\D(\u^{\pm})\Id + 2\mu^{\pm} \D(\u^{\pm}), \quad 
  \tr\D(\u^{\pm}) = \div\u^\pm,
\end{equation}
where $\Id := (\delta_{ij})_{1\leq i,j\leq 3}$ is the identity tensor, $\D(\u) := \dfrac{1}{2}\Bigl(\nabla\u^{\pm} + {}^t(\nabla\u^{\pm})\Bigr)$ is the deformation rate and $\lambda^{\pm}$, $\mu^{\pm}$ are viscosity coefficients. For ideal gases, for example, these coefficients are related by Stokes relation $\lambda^\pm + \frac23\mu^\pm = 0$. In application to powder-snow avalanches, viscosity coefficients $\lambda^{\pm}$, $\mu^{\pm}$ should be understood in the sense of eddy viscosity.

\begin{remark}
From physical point of view, presented here model (\ref{eq:mass4}), (\ref{eq:momentum4}) is far from being complete. For example, one could supplement it by capillarity effects in the Korteweg form. Also we omited all the terms which model mass, momentum and energy exchange between two phases. Generally, their form is strongly dependent on the physical situation under consideration.
\end{remark}

\begin{remark}
Since we do not consider the total energy conservation equation, the fluids are implicitly assumed to be barotropic. In the absence of viscous stresses $\ttau^\pm$, the flow is isentropic. This simplification can be adopted provided that important energy transfers do not occur. Non-isentropic flows are considered in \cite{Meyapin2009}.
\end{remark}

\begin{remark}\label{rem:massf}
While considering two-phase flows, it is useful to introduce several additional quantities which play an important r\^ole in the description of such flows. The mixture density $\rho$ and mass fractions $m^\pm$ are naturally defined as:
\begin{equation*}
  \rho (\x,t) := \alpha^+\rho^+ + \alpha^-\rho^- > 0, \quad \forall (\x,t) \in \Omega\times [0, T], 
\end{equation*}
\begin{equation*}
  m^\pm := \frac{\alpha^\pm\rho^\pm}{\rho}, \quad m^+ + m^- = 1.
\end{equation*}
The total density $\rho$ is assumed to be strictly positive everywhere in the domain $\Omega$. Hence, the void creation is forbidden in our modeling. 

Important quantities $\rho$, $m^\pm$ will appear several times below.
\end{remark}

In principle, one could use equations (\ref{eq:mass4}), (\ref{eq:momentum4}) to model various two-phase flows. However, this system remains quite expensive for large scale simulations required by real-life applications. The major difficulty comes from the advection operator associated to model (\ref{eq:mass4}), (\ref{eq:momentum4}) which can be non-hyperbolic \cite{Bresch2008, Rovarch2006}. In the next section we will derive a simplified two-fluid model which is proposed as a candidate for powder-snow avalanche compressible simulations.

\section{Velocity relaxation}\label{sec:relax}

We would like to reduce the number of variables in the system (\ref{eq:mass4}), (\ref{eq:momentum4}). The main idea is to introduce the common velocity field for both phases. For this purpose, we will introduce a relaxation term to the momentum conservation equation (\ref{eq:momentum4}):
\begin{equation}\label{eq:momentumRelax}
  \dt(\alpha^{\pm}\rho^{\pm}\u^{\pm}) + \div(\alpha^{\pm}\rho^{\pm}\u^{\pm}\otimes\u^{\pm}) + 
	\alpha^{\pm}\nabla p = \div(\alpha^{\pm}\ttau^{\pm})+\alpha^{\pm}\rho^{\pm}\g
	\pm\dfrac{\kappa}{\epsilon}(\u^+ - \u^-),
\end{equation}
where $\kappa = \O(1)$ is a constant and $\epsilon$ is a small parameter which controls the magnitude of the relaxation term. Physically this additional term represents the friction between two phases. In the following, we are going to take the singular limit as the relaxation parameter $\epsilon\to 0$. This is achieved with Chapman-Enskog type expansion. In this way, we constrain velocities $\u^\pm (\x,t)$ to tend to the common value $\u (\x,t)$. This technique has been already successfully applied to the Baer-Nunziato model \cite{Baer1986} in \cite{Murrone2005}.

The first step consists in rewriting the governing equations (\ref{eq:mass4}), (\ref{eq:momentumRelax}) in the quasilinear form. To shorten notations, we will also use the material time derivative which is classically defined for any smooth scalar function $\phi (\x,t)$ as
\begin{equation*}
  \od{^\pm\phi}{t} := \pd{\phi}{t} + \u^\pm \cdot \nabla\phi.
\end{equation*}

\begin{lemma}
Smooth solutions to equations (\ref{eq:mass4}), (\ref{eq:momentumRelax}) satisfy the following system:
\begin{eqnarray}\label{eq:pressDt}
  \alpha^\pm\od{^\pm p}{t} + \rho^\pm(c_s^\pm)^2\od{^\pm \alpha^\pm}{t} + 
  \alpha^\pm \rho^\pm(c_s^\pm)^2 \div\u^\pm = 0, \\
  \alpha^\pm\rho^\pm\od{^\pm\u^\pm}{t} + \alpha^\pm\nabla p = 
  \div (\alpha^\pm\ttau^\pm) + \alpha^\pm\rho^\pm\g
  \pm\frac{\kappa}{\epsilon}(\u^+ - \u^-), \label{eq:uDt}
\end{eqnarray}
where $(c_s^\pm)^2 := \left.\pd{p^\pm}{\rho^\pm}\right|_{s^\pm}$ represents the sound speed in each phase $\pm$.
\end{lemma}

\begin{proof}
This result follows from direct calculations. First of all, we remark that the mass conservation equation (\ref{eq:mass4}) can be rewritten using the material derivative as follows:
\begin{equation}\label{eq:massDt}
\od{^\pm(\alpha^\pm\rho^\pm)}{t} + \alpha^\pm\rho^\pm\div\u^\pm = 0.
\end{equation} 
Using equations of state $p = p^\pm(\rho^\pm)$, we can express the density material derivative in terms of the pressure and the sound speed:
\begin{equation*}
  \od{^\pm \rho^\pm}{t} = \frac{1}{(c_s^\pm)^2}\od{^\pm p}{t}.
\end{equation*}
Now, it is straightforward to derive equation (\ref{eq:pressDt}) from (\ref{eq:massDt}).

Finally, if we multiply equation (\ref{eq:massDt}) by $\u^\pm$ and subtract it from the momentum conservation equation (\ref{eq:momentumRelax}), we will get desired result (\ref{eq:uDt}).
\end{proof}

Equations (\ref{eq:pressDt}), (\ref{eq:uDt}) can be also recast in the matrix form which is particularly useful for further developments:
\begin{equation}\label{eq:matrix}
  \A(\Ve)\pd{\Ve}{t} + \B(\Ve)\nabla\Ve = \div\T(\Ve) + \S(\Ve) + \frac{\Rr(\Ve)}{\epsilon},
\end{equation}
where we introduced several notations. The vector $\Ve$ represents four unknown physical variables $\Ve := {}^t (p, \alpha^+, \u^+, \u^-)$ and $\pd{\Ve}{t} := {}^t(\dt p, \dt\alpha^+, \dt\u^+, \dt\u^-)$ and $\nabla\Ve := {}^t\bigl(\nabla p, \nabla\alpha^+, (\cdot\nabla)\u^+, (\cdot\nabla)\u^-\bigr)$. Matrices $\A(\Ve)$ and $\B(\Ve)$ are defined as
\begin{equation*}
  \A(\Ve) := \begin{pmatrix}
  				  \alpha^+ & \rho^+(c_s^+)^2 & 0 & 0 \\
  				  \alpha^- & -\rho^-(c_s^-)^2 & 0 & 0 \\
  				  0 & 0 & \alpha^+\rho^+\Id & 0 \\
  				  0 & 0 & 0 & \alpha^-\rho^-\Id \\
  				\end{pmatrix},
\end{equation*}
\begin{equation*}
  \B(\Ve) := \begin{pmatrix}
  					\alpha^+\u^+ & \rho^+(c_s^+)^2\u^+ & \alpha^+\rho^+(c_s^+)^2\Id & 0 \\
  					\alpha^-\u^- & -\rho^-(c_s^-)^2\u^- & 0 & \alpha^-\rho^-(c_s^-)^2\Id \\
  					\alpha^+\Id & 0 & \alpha^+\rho^+\u^+ & 0 \\
  					0 & \alpha^-\Id & 0 & \alpha^-\rho^-\u^- \\
  				 \end{pmatrix}.
\end{equation*}
In these matrix notations the size of zero entries must be chosen to make the multiplication operation possible.

On the right hand side of (\ref{eq:matrix}), the work of viscous forces is denoted by symbol $\div\T(\Ve) := {}^t(0, 0, \div\ttau^+, \div\ttau^-)$. The source term $\S(\Ve) := {}^t(0, 0, \alpha^+\rho^+\g, \alpha^-\rho^-\g)$ incorporates the gravity force and $\Rr(\Ve) := {}^t(0, 0, \kappa(\u^+ - \u^-), -\kappa(\u^+ - \u^-))$ contains the relaxation terms.

Since we expect the limit $\Ve\to V$ to be finite as $\epsilon\to 0$, necessary the limiting vector $V$ lies in the hypersurface $\Rr(V) = 0$. In terms of physical variables, it implies $\u^+ \equiv \u^-$. Consequently, we find our solution in the form of the following Chapman-Enskog type expansion:
\begin{equation*}
  \Ve = V + \epsilon W + \O(\epsilon^2).
\end{equation*}
After substituting this expansion into (\ref{eq:matrix}) and taking into account that $\Rr(V) \equiv 0$, at the leading order in $\epsilon$ one obtains:
\begin{equation}\label{eq:asympt}
  \A(V)\pd{V}{t} + \B(V)\nabla V = \div\T(V) + \S(V) + \Rr'(V)W,
\end{equation}
where 
\begin{equation*}
  \Rr'(V) := \begin{pmatrix}
   					  0 & 0 & 0 & 0 \\
   					  0 & 0 & 0 & 0 \\
   					  0 & 0 & \kappa\Id & -\kappa\Id \\
   					  0 & 0 & -\kappa\Id & \kappa\Id \\
  					 \end{pmatrix}
\end{equation*}

Henceforth, we make a technical assumption of the presence of both phases in any point $\x\in\Omega$ of the flow domain. Mathematically it means that $0 < \alpha^+ < 1$. Since $\alpha^+ + \alpha^- = 1$, the same inequality holds for $\alpha^-$. Otherwise, the relaxation process physically does not make sense and we will have some mathematical technical difficulties.

Under the aforementioned assumption, the matrix $\A(V)$ is invertible. Hence, we can multiply on the left both sides of (\ref{eq:asympt}) by $\P\A^{-1}(V)$ where the projection matrix $\P$ is to be specified below:
\begin{equation}\label{eq:Peq}
  \P\pd{V}{t} + \P\A^{-1}(V)\B(V)\nabla V = \P\A^{-1}(V)\div\T(V) + \P\tilde\Rr'(V)W + \P\A^{-1}(V)\S(V),
\end{equation}
where $\tilde\Rr'(V) := \A^{-1}(V)\Rr'(V)$ and has the following components
\begin{equation*}
  \tilde\Rr'(V) = \begin{pmatrix}
  								  0 & 0 & 0 & 0 \\
  								  0 & 0 & 0 & 0 \\
  								  0 & 0 & \displaystyle\frac{\kappa}{\alpha^+\rho^+}\Id & -\displaystyle\frac{\kappa}{\alpha^+\rho^+}\Id \\
  								  0 & 0 & -\displaystyle\frac{\kappa}{\alpha^-\rho^-}\Id & \displaystyle\frac{\kappa}{\alpha^-\rho^-}\Id \\
  								\end{pmatrix}.
\end{equation*}

The vector of physical variables $V$ has four (in 1D) components ${}^t(p, \alpha^+, \u, \u)$ and only three are different. In order to remove the redundant information, we will introduce the new vector $U$ defined as $U := {}^t(p, \alpha^+, \u)$. The Jacobian matrix of this transformation can be easily computed:
\begin{equation*}
  \J := \pd{V}{U} = \begin{pmatrix}
  										1 & 0 & 0 \\
  										0 & 1 & 0 \\
  										0 & 0 & \Id \\
  										0 & 0 & \Id \\
  									\end{pmatrix}.
\end{equation*}
In new variables equation (\ref{eq:Peq}) becomes:
\begin{equation}\label{eq:Peq2}
	\P\J\pd{U}{t} + \P\A^{-1}(U)\B(U)\J\nabla U = \P\A^{-1}(U)\div\T(U) + \P\tilde\Rr'(U)W + \P\A^{-1}(U)\S(U).
\end{equation}

Now we can formulate two conditions to construct the matrix $\P$. First of all, the vector $W$ is unknown and we need to remove it from equation (\ref{eq:Peq2}). Hence, we require $\P\tilde\Rr'(V) = 0$. Then, we would like the governing equations to be explicitly resolved with respect to time derivatives. It gives us the second condition $\P\J = \Id$. The existence and effective construction of the matrix $\P$ satisfying two aforementioned conditions
\begin{equation*}
  \P\tilde\Rr'(V) = 0, \quad \P\J = \Id,
\end{equation*}
are discussed below. Presented in this section results follow in great lines \cite{Murrone2005}.

We will consider a slightly more general setting. Let vector $V\in\R^n$ and its reduced counterpart $U\in\R^{n-k}$, $k < n$. In such geometry, $\tilde\Rr'(V)\in\Mat_{n,n}(\R)$, $\J\in\Mat_{n,n-k}(\R)$ and, consequently, $\P\in\Mat_{n-k,n} (\R)$. Here, the notation $\Mat_{m,n}(\R)$ denotes the set of $m \times n$ matrices with coefficients in $\R$. We have to say also that from algebraic point of view, matrices $\tilde\Rr'(V)$ and $\Rr'(V)$ are completely equivalent. Thus, for simplicity, in the following propositions we will reason in terms of $\Rr'(V)$.

\begin{lemma}\label{lemma:J}
The columns of the Jacobian matrix $\J$ form a basis of $\ker\bigl(\Rr'(V)\bigr)$.
\end{lemma}
\begin{proof}
If we differentiate the relation $\Rr(V) = 0$ with respect to $U$, we will get the identity $\Rr'(V)\J = 0$. It implies that $\range\bigl(\J\bigr) \subseteq \ker \bigl(\Rr'(V)\bigr)$. By direct computation one verifies that $\dim\range\bigl(\Rr'(V)\bigr) = k$. From the well-known identity $\range\bigl(\Rr'(V)\bigr)\oplus \ker\bigl(\Rr'(V)\bigr) = \R^n$, one concludes that $\dim\ker\bigl(\Rr'(V)\bigr) = n-k$. But in the same time, the rank of $\J$ is equal to $n-k$ as well. It proves the result.
\end{proof}

\begin{theorem}
We suppose that for all $V$, $\range\bigl(\Rr'(V)\bigr) \cap \ker\bigl(\Rr'(V)\bigr) = \{0\}$ then it exists a matrix $\P\in\Mat_{n-k,n} (\R)$ such that $\P\Rr'(V) = 0$ and $\P\J = \Id_{n-k}$.
\end{theorem}
\begin{proof}
Hypothesis $\range\bigl(\Rr'(V)\bigr) \cap \ker\bigl(\Rr'(V)\bigr) = \{0\}$ implies that $\range\bigl(\Rr'(V)\bigr)\oplus \ker\bigl(\Rr'(V)\bigr) = \R^n$. From Lemma \ref{lemma:J} it follows that $\range\bigl(\J\bigr) = \ker\bigl(\Rr'(V)\bigr)$. Thus, the space $\R^n$ can be also represented as a direct sum $\range\bigl(\Rr'(V)\bigr)\oplus \range\bigl(\J\bigr)$. We will define $\P$ to be the projection on $\ker\bigl(\Rr'(V)\bigr) \equiv \range\bigl(\J\bigr)$. Since obviously $\Rr'(V)\in \range\bigl(\Rr'(V)\bigr)$ and $\J\in\range\bigl(\J\bigr)$, we have two required identities: $\P\J = \Id_{n-k}$ and $\P\Rr'(V) = 0$.
\end{proof}

Now, in order to compute effectively the projection matrix $\P$, we will construct an auxiliary matrix $\Dv(V) = [J^1,\ldots, J^{n-k}, I^1,\ldots,I^k]$, where $J^i$ is the column $i$ of the matrix $\J$ and $\{I^1,\ldots,I^k\}$ are vectors which form a basis of $\range\bigl(\Rr'(V)\bigr)$. We remark that $\P\Dv(V) = [\Id_{n-k}, 0]$. Lemma \ref{lemma:J} implies that the matrix $\Dv(V)$ is invertible. Thus, the projection $\P$ can be computed by inverting $\Dv(V)$:
\begin{equation*}
  \P = [\Id_{n-k}, 0]\cdot\Dv^{-1}(V).
\end{equation*}

Let us apply this general framework to our model (\ref{eq:Peq}), where $n = 4$ and $k = 1$. The matrix $\Dv(V)$ and its inverse $\Dv^{-1}(V)$ take this form:
\begin{equation*}
  \Dv(V) = \begin{pmatrix}
  					1 & 0 & 0 & 0 \\
  					0 & 1 & 0 & 0 \\
  					0 & 0 & \Id & \displaystyle\frac{\kappa}{\alpha^+\rho^+}\Id \\
  					0 & 0 & \Id & -\displaystyle\frac{\kappa}{\alpha^-\rho^-}\Id \\
  				 \end{pmatrix}, \quad
  \Dv^{-1}(V) = \begin{pmatrix}
  							 1 & 0 & 0 & 0 \\
  							 0 & 1 & 0 & 0 \\
  							 0 & 0 & m^+\Id & m^-\Id \\
  							 0 & 0 & \displaystyle\frac{m^+m^-\rho}{\kappa}\Id & -\displaystyle\frac{m^+m^-\rho}{\kappa}\Id \\
  							\end{pmatrix},
\end{equation*}
where $m^\pm$ are mass fractions defined in Remark \ref{rem:massf}.

Now, the projection matrix $\P$ can be immediately computed:
\begin{equation*}
  \P = \begin{pmatrix}
  		  1 & 0 & 0 & 0 \\
  		  0 & 1 & 0 & 0 \\
  		  0 & 0 & m^+\Id & m^-\Id \\
  		 \end{pmatrix}.
\end{equation*}
Finally, after computing all matrix products $\P\A^{-1}(U)\B(U)\J$, $\P\A^{-1}(U)\div\T(U)$, $\P\A^{-1}(U)\S(U)$ present in equation (\ref{eq:Peq2}), we obtain the desired single velocity model:
\begin{eqnarray}\label{eq:first}
  \pd{p}{t} + \u\cdot\nabla p + \rho c_s^2\div\u = 0, \\
  \pd{\alpha^+}{t} + \u\cdot\nabla\alpha^+ + \alpha^+\alpha^-\delta\div\u = 0, \label{eq:second} \\
  \rho\pd{\u}{t} + \rho(\u\cdot\nabla)\u = \rho\g + \div\ttau, \label{eq:last}
\end{eqnarray}
where $\rho = \alpha^+\rho^+ + \alpha^-\rho^-$ is the mixture density and $c_s^2$ is the sound velocity in the mixture which is determined by this formula:
\begin{equation*}
  \rho c_s^2 := \frac{\rho^+\rho^-(c_s^+)^2(c_s^-)^2}{\alpha^-\rho^+(c_s^+)^2 + \alpha^+\rho^-(c_s^-)^2},
\end{equation*}
and $\delta$ is given by
\begin{equation*}
  \delta := \frac{\rho^+(c_s^+)^2 - \rho^-(c_s^-)^2}{\alpha^-\rho^+(c_s^+)^2 + \alpha^+\rho^-(c_s^-)^2}.
\end{equation*}
Finally, $\ttau := \lambda\tr\D(\u)\Id + 2\mu\D(\u)$ is the viscous stress tensor of the mixture. Viscosity coefficients $\lambda$, $\mu$ are naturally defined as
\begin{equation*}
  \lambda := \alpha^+\lambda^+ + \alpha^-\lambda^-, \quad
  \mu := \alpha^+\mu^+ + \alpha^-\mu^-.
\end{equation*}

Equations (\ref{eq:first}) -- (\ref{eq:last}) can be recast in the conservative form which is more convenient for numerical computations and theoretical analysis. To achieve this purpose, we replace the pressure $p$ in (\ref{eq:first}) by $\rho^\pm$ using the equation of state: 
\begin{equation*}
  \pd{\rho^\pm}{t} + \u\cdot\nabla\rho^\pm + \frac{\rho c_s^2}{(c_s^\pm)^2}\div\u = 0.
\end{equation*}
The last equation is multiplied by $\alpha^\pm$, the second equation (\ref{eq:second}) is multiplied by $\rho^\pm$ and we sum them to come up with two mass conservation equations. Transformation of the momentum conservation equation (\ref{eq:last}) is straightforward. The resulting conservative system takes this form:
\begin{eqnarray}\label{eq:massCons}
 \dt(\alpha^\pm\rho^\pm) + \div(\alpha^\pm\rho^\pm\u) = 0, \\
 \dt(\rho\u) + \div(\rho\u\otimes\u) + \nabla p = \div\ttau + \rho\g. \label{eq:momentumCons}
\end{eqnarray}
These equations represent a barotropic version of the four-equations model proposed in \cite{Dias2008, Dutykh2007a}.

It can be shown that the advection operator of the model (\ref{eq:massCons}), (\ref{eq:momentumCons}) is hyperbolic for any reasonable equation of state (\ref{eq:EOS}). Moreover, this system contains fewer variables which allow more efficient computations required in practice.

\section{Incompressible limit}\label{sec:lowMach}

The main scope of this paper is certainly around compressible two-fluid models. However, we decided to derive an incompressible limit of the single velocity model (\ref{eq:massCons}), (\ref{eq:momentumCons}) for the case when acoustic effects should be filtered out. The presence of acoustic waves represent, for example, a major restriction for the time step, if an explicit scheme is used. 

For the sake of simplicity, we will neglect dissipative effects which do not affect the acoustic wave propagation. Thus, in this section we consider the following system of equations:
\begin{eqnarray}\label{eq:mass2}
  \dt(\alpha^\pm\rho^\pm) + \div(\alpha^\pm\rho^\pm\u) = 0, \\
  \rho\dt\u + \rho(\u\cdot\nabla)\u + \nabla p = \rho\g. \label{eq:mom2} \\ 
\end{eqnarray}
For convenience, we rewrite equation (\ref{eq:momentumCons}) in nonconservative form.

In order to estimate the relative importance of various terms, we introduce dimensionless variables. The characteristic length, time, and velocity scales are denoted by $\ell$, $t_0$ and $U_0$ respectively. For example, $\ell$ may be chosen as the diameter of the fluid domain $\Omega$, $t_0$ is  the biggest vortex turnover time and $U_0$ is the typical flow velocity. The density and the sound velocity scales are chosen to be those of the heavy fluid, i.e. $\rho_0^+$ and $c_{0s}^+$ correspondingly. Since we are interested in acoustic effects, the natural pressure scale is given by $\rho_0^+(c_{0s}^+)^2$. If we summarize these remarks, dependent and independent dimensionless variables (denoted with primes) are defined as:
\begin{equation*}
  \x' := \frac{\x}{\ell}, \quad t' := \frac{t}{t_0}, \quad \u' := \frac{\u}{U_0}, \quad 
  (\rho^\pm)' := \frac{\rho^\pm}{\rho_0^+}, \quad p' := \frac{p}{\rho_0^+(c_{0s}^+)^2}.
\end{equation*}

\begin{remark}
There is nothing to do for the volume fractions $\alpha^\pm$, since this quantity is dimensionless by definition.
\end{remark}

After dropping the tildes, nondimensional system of equation becomes:
\begin{eqnarray}\label{eq:mass3}
  \St\dt(\alpha^\pm\rho^\pm) + \div(\alpha^\pm\rho^\pm\u) = 0, \\
  \St\rho\dt\u + \rho(\u\cdot\nabla)\u + \frac{1}{\Ma^2}\nabla p = \frac{1}{\Fr^2}\rho\g, \label{eq:mom3}
\end{eqnarray}
where several scaling parameters have appeared:
\begin{itemize}
  \item Strouhal number $\St := \displaystyle\frac{\ell}{U_0 t_0}$. In this study we will assume the Strouhal number to be equal to one, i.e. $t_0 = \displaystyle\frac{\ell}{U_0}$.
  \item Mach number $\Ma := \displaystyle\frac{U_0}{c_{0s}^+}$ which measures the relative importance of the flow speed and the sound speed in the medium.
  \item Froude number $\Fr := \displaystyle\frac{U_0}{\sqrt{g\ell}}$ compares inertia and gravitational force. This parameter will not play an important r\^ole in the present study.
\end{itemize}
All physical variables $\alpha^\pm$, $\rho^\pm$, $p$ and $\u$ are expanded in formal series in powers of the Mach number:
\begin{equation}\label{eq:expansion}
  \phi = \phi_0 + \Ma\phi_1 + \Ma^2\phi_2 + \ldots, \qquad
  \phi \in \{\alpha^\pm, \rho^\pm, p, \u \}.
\end{equation}
Formal expansion (\ref{eq:expansion}) is then substituted into the system (\ref{eq:mass3}), (\ref{eq:mom3}). At the orders $\Ma^{-2}$ and $\Ma^{-1}$, we obtain
\begin{equation*}
  \nabla p_0 = \nabla p_1 = 0.
\end{equation*}
In other words, $p_0 = p_0 (t)$ and $p_1 = p_1(t)$ are only functions of time. At the order $\Ma^0$ we get the following system of equations:
\begin{eqnarray}\label{eq:mass0}
  \dt(\alpha_0^\pm\rho_0^\pm) + \div(\alpha_0^\pm\rho_0^\pm\u_0) = 0, \\
  \rho_0\dt\u_0 + \rho_0(\u_0\cdot\nabla)\u_0 + \nabla\pi = \frac{1}{\Fr^2}\rho_0\g, \\
\end{eqnarray}
where by $\pi$ we denote $p_2$.

Using the same asymptotic expansion (\ref{eq:expansion}), one can show that at the leading order we keep usual relations between densities and volume fractions:
\begin{equation}\label{eq:alpha0}
  \alpha_0^+ + \alpha_0^- = 1, \quad \rho_0 = \alpha_0^+\rho_0^+ + \alpha_0^-\rho_0^-.
\end{equation}
In order to investigate the behaviour of $\rho_0^\pm$, we will invert the equation of state\footnote{The function $p = p^\pm(\rho^\pm)$ is invertible since it is a strictly increasing function $\displaystyle\pd{p}{\rho^\pm} > 0$.} $\rho^\pm = \rho^\pm (p) = (p^\pm)^{-1} (p)$ and expand it in powers of $\Ma$:
\begin{equation*}
  \rho^\pm(p) = \rho^\pm (p_0) + \Ma\left.\pd{\rho^\pm}{p}\right|_{p_0}p_1 + 
  \Ma^2 \Bigl(\left.\pd{\rho^\pm}{p}\right|_{p_0}p_2 + 
  \left.\pd{^2\rho^\pm}{p^2}\right|_{p_0}p_1^2\Bigr) + \O (\Ma^3)
\end{equation*}
On the other hand, from (\ref{eq:expansion}) we know that
\begin{equation*}
	\rho^\pm = \rho_0^\pm + \Ma\rho_1^\pm + \Ma^2\rho_2^\pm + \ldots
\end{equation*}
Matching these expansions at two lowest orders shows that $\rho_{0,1}^\pm$ are functions only of the time variable:
\begin{equation*}
  \rho_0^\pm = \rho^\pm (p_0 (t)) =: r_0^\pm (t), \quad
  \rho_1^\pm = \left.\pd{\rho^\pm}{p}\right|_{p_0(t)}p_1(t) =: r_1^\pm (t).
\end{equation*}
It is possible to show that $\rho_{0,1}^\pm$ are just constants. Consider the Gibbs relation which reads
\begin{equation*}
  T^\pm ds^\pm = de^\pm - \frac{p}{(\rho^\pm)^2}d\rho^\pm.
\end{equation*}
Since we consider isentropic flows, $ds^\pm \equiv 0$ and, consequently, the Gibbs relation takes a much simpler form:
\begin{equation}\label{eq:gibbs}
  de^\pm = \frac{p}{(\rho^\pm)^2}d\rho^\pm.
\end{equation}
It can be shown by considering the total energy conservation equation \cite{Meyapin2009}, that the internal energy $e^\pm$ naturally scales with $U_0^2$. After dividing (\ref{eq:gibbs}) by $dt$ and switching to dimensionless variables, equation (\ref{eq:gibbs}) takes the following form (after droping the primes):
\begin{equation*}
  \od{e^\pm}{t} = \frac{p}{\Ma^2 (\rho^\pm)^2}\od{\rho^\pm}{t}.
\end{equation*}
Expanding $e^\pm$ in the series (\ref{eq:expansion}) and looking at two leading terms, leads to the desired result:
\begin{equation*}
  \od{\rho_{0,1}^\pm}{t} = 0 \quad \Rightarrow \quad \rho_{0,1}^\pm = \mathrm{const}.
\end{equation*}
The incompressibility condition $\div\u_0 = 0$ is obtained by summing up mass conservation equations (\ref{eq:mass0}) and taking into account relation (\ref{eq:alpha0}).

If we summarize all developments made above and switch back to dimensional variables, the resulting incompressible system will become:
\begin{eqnarray}\label{eq:ns1}
	\dt\alpha^\pm + \nabla\alpha^\pm\cdot\u = 0, \\
	\div\u = 0, \label{eq:ns2} \\
  \rho\dt\u + \rho(\u\cdot\nabla)\u + \nabla\pi = \rho\g + \div\ttau, \label{eq:ns3}
\end{eqnarray}
where we dropped the index $0$ and added again dissipative effects. Viscous stress tensor $\ttau$ is still defined by expression (\ref{eq:viscous}), as in compressible case. In this case, we can speak about two-fluid Navier-Stokes equations. This system of equations (\ref{eq:ns1}) -- (\ref{eq:ns3}) is much easier to solve numerically than its compressible analogue (\ref{eq:massCons}), (\ref{eq:momentumCons}). In particular, this simplification is due to removed stiffness of acoustic waves.

\section{Conclusions and perspectives}\label{sec:concl}

In this study we presented several barotropic two-fluid models which can be used for numerical simulation of powder-snow avalanche flows. One of the main objectives of this paper was to reveal the connection between barotropic models with single and two velocities. The extension to more general fluids is in progress \cite{Meyapin2009}.

Our exposition began with compressible two-phase model (\ref{eq:mass4}), (\ref{eq:momentum4}) possessing two velocity variables. Then, using a relaxation process, we constrained the system to have a common velocity for both phases. Mathematically it was achieved with a Chapman-Enskog type expansion. Resulting model (\ref{eq:massCons}), (\ref{eq:momentumCons}) is hyperbolic for any reasonable equation of state (\ref{eq:EOS}). Finally, two-fluid Navier-Stokes equations (\ref{eq:ns1}) -- (\ref{eq:ns3}) were derived as an incompressible limit of the single velocity model (\ref{eq:massCons}), (\ref{eq:momentumCons}). 

Hence, we presented three different two-fluid models which are related by formal derivation procedures. Simplifications made above, represent a good trade-off between accuracy and computational complexity. The final choice should be made after determining the flow r\'egime and main goals of the simulation.

We did not incorporate yet any turbulence modeling. In this study we were focused essentially on the advection operators. However, it is obvious that the physical flow under consideration is fully turbulent in its aerosol part \cite{Rastello2004}. As the first physical approximation, turbulence effects can be taken into account by adding eddy viscosity terms and, thus, by modifying the viscous stress tensor $\ttau$. It will be done in future studies.

\section*{Acknowledgement}

The authors would like to acknowledge the University of Savoie for the PPF grant linked to the project: ``Math\'ematiques et avalanches de neige, une rencontre possible?''. The support from the Research network VOR (Professors Jacky Mazars and Denis Jongmans) and Cluster Environnement through the program ``Risques gravitaires, séismes'' is also acknowledged.

We would like to thank Professor Carmen de Jong for interesting discussions around snow avalanches. Special thanks go to our colleagues and friends Didier Bresch and C\'eline Acary-Robert for their continuous help and support. Finally, the second author thanks Professors Jean-Michel Ghidaglia and Fr\'ed\'eric Dias for introducing him to the beautiful field of two-phase flows.

\bibliographystyle{spmpsci}
\bibliography{biblio}

\end{document}